\documentstyle[preprint,aps]{revtex}

\begin{document}
\thispagestyle{empty}
\begin{titlepage}
\draft
\preprint{CAS-HEP-T-95-06/001;  OHSTPY-HEP-T-95-016}
\title{A Predictive Superunification
$SO(10)\times \Delta_{48}(SU(3))\times U(1)$ Model for
Fermion Masses and Mixings}
\author{K.C. Chou$^{1}$ and Y.L. Wu$^{2}$\footnote{
supported in part by Department of Energy Grant\# DOE/ER/01545-655}}
\address{ $^{1}$Chinese Academy of Sciences, Beijing 100864,  China
 \\ $^{2}$Department of Physics, \ Ohio State  University \\ Columbus,
 Ohio 43210,\ U.S.A. }
\date{June 1995, hep-ph/9506312}
\maketitle

\begin{abstract}
 Based on a few basic assumptions, a predictive supersymmetric grand
unification $G_{U}=SO(10)\times \Delta_{48}(SU(3))\times U(1)$ model
is proposed. Thus, all the observed 45
chiral fermions with additional 3 right-handed neutrinos are unified into an
irreducible triplet 16-dimensional spinor representation of $G_{U}$.
Electroweak symmetry breaking
naturally requires two Higgs doublets which belong to a single 10
representation of $SO(10)$. The model provides an unified description for all
the Yukawa and gauge couplings. With a given structure of the
physical vacuum, the fermion masses and mixings are predicted
in terms of two basic parameters.
\end{abstract}

\end{titlepage}

\narrowtext


   A great success of the standard model (SM) motivates us to explore a
more fundamental theory to understand the origin of the eighteen parameters
in the SM.  Large efforts have been made in this field.
The hope of this note is to systematize the
most attractive ideas and motivations that, we think, nature might choose,
to constitute the basic assumptions for building
a realistic model with a minimum set of parameters.
Our starting point is based on the following basic assumptions

{\bf i)}\  All fields of the fundamental particles
have definite parity. These fields
consist of matter fields with spin $s=1/2$, which constitute the basic building
blocks of our universe; Yang-Mills gauge fields with spin $s=1$,
which describe the basic forces of nature and glue the basic
building blocks together to form our universe;
and scalar fields with spin $s=0$ which provide the symmetry-breaking force and
give the boundary conditions of our universe as well as determine the structure
of the physical vacuum.  All gauge interactions,  (the known ones are
the electromagnetic, weak, and strong interactions with a
 gauge symmetry $G_{SM}= U(1)_Y \times
SU(2)_{L} \times SU(3)_{c}$), are assumed to be unified at a high energy scale.

 {\bf ii)}\  The theory should be  invariant under
the gauge as well as Lorentz transformations and be also anomaly free.
 Physical laws are left-right symmetric and time-reversal invariant before
the unified-gauge-symmetry breaking. The missing symmetries
result from the physical vacuum via spontaneous symmetry breaking and/or
dynamical symmetry breaking.

 {\bf iii)}\   All matter fields (the known ones are the three families of
the quarks and leptons) should be treated on the same footing at an energy
scale $\bar{M}_{P}$ which is close to the
Planck scale $M_{P}$,  namely all matter fields should be democratic
and one can not tell which quark or
lepton it is  and to which family it belongs at that high energy
scale. The same considerations should also be applied to the
physical Higgs scalar fields.

 {\bf iv)}\  Nature is supersymmetric. The
supersymmetry is expected to be broken at
an energy  scale $M_{S}\sim $TeV.

     From the first assumption i),  a minimal grand unification theory (GUT)
 which unifies the known gauge interactions
$G_{SM} = U(1)_{Y} \times SU(2)_{L} \times SU(3)_{c}$ is SU(5).
But it does not satisfy the  second assumption ii). The second assupmtion
implies that the left-handed and right-handed fermions should be
symmetrically treated.  It is not difficult to show that SO(10)
is a minimal GUT which can simultaneously satisfy the assumptions i) and ii).
This is because SO(10) contains $SU(2)_L \times SU(2)_R$ symmetry
which is obviously left-right symmetric.
Thus, the fifteen chiral fermions of each
family with an additional right-handed
neutrino are unified into an
irreducible 16-dimensional spinor representation of $SO(10)$.

   From the third assumption iii),  the three families
should also be unified into
 an irreducible triplet representation of  a family symmetry group.
Based on the known fact that fermions in the third family are much heavier
than those in the first and second families.  It implies that the
family symmetry group must be a group with at
least rank three. However, within the known simple continuous groups, there is
a
difficulty to find a rank three group which has irreducible triplet
representations. This limitation of the continuous groups
can be avoided by considering their finite and disconnected subgroups.
  A simple and an interesting example is the finite and disconnected subgroups
of SU(3), which contain an arbitrary number of irreducible triplet
representations.  Of particular interesting subgroup is
the $\Delta_{3n^{2}}(SU(3))$  dihedral subgroups\cite{SUBGROUP} of SU(3) since
their irreducible representations contain solely triplets and singlets.
Here we will present a model based on
$\Delta_{48}(SU(3))$ (i.e., n=4 ) which is the smallest of the dihedral
group $\Delta_{3n^{2}}(SU(3))$  with $n/3$ being not an interger.

   The irreducible triplet representations of $\Delta_{48}(SU(3))$ consist of
two complex triplets $T_{1}(\bar{T}_{1})$ and $T_{3}(\bar{T}_{3})$ and one real
triplet $T_{2} = \bar{T}_{2}$ as well as three singlet representations.
Their irreducible triplet representations can be expressed
in terms of the matrix representation
\begin{eqnarray}
T_{1}^{(1)} & = & diag. (i, 1, -i), \qquad T_{1}^{(2)}  = diag. (1, -i, i),
\qquad T_{1}^{(3)}  = diag. (-i, i, 1); \nonumber  \\
T_{2}^{(1)} & = & diag. (-1, 1, -1), \qquad T_{2}^{(2)}  = diag. (1, -1, -1),
\qquad T_{2}^{(3)}  = diag. (-1, -1, 1);  \\
T_{3}^{(1)} & = & diag. (i, -1, i), \qquad T_{3}^{(2)}  = diag. (-1, i, i),
\qquad T_{3}^{(3)}  = diag. (i, i, -1) \nonumber
\end{eqnarray}
The matrix representations of $\bar{T}_{1}^{(i)}$ and $\bar{T}_{3}^{(i)}$ are
the hermician conjugates of $T_{1}^{(i)}$ and $T_{3}^{(i)}$.
{}From this representation, we can explicitly construct the invariant tensors.

   With the above considerations, all three families with $3\times 16 = 48$
chiral fermions can be unified into a triplet 16-dimensional spinor
representation of $SO(10)\times \Delta_{48}(SU(3))$.
Without losing generality, one can assign the three
chiral families into the triplet representation $T_{1}$, which may be simply
denoted as $\hat{16} = 16_{i} T_{1}^{(i)}$.
All the fermions are assumed to obtain their masses through a single
$10_{1}$ of SO(10) into which the needed two Higgs doublets are unified.
  In comparison with the SM, SO(10) model naturally  requires
two Higgs doublets\footnote{A general two-Higgs doublet model (2HDM) motivated
from CP violation and superweak interaction see, Y.L. Wu and L. Wolfenstein,
Phys. Rev. Lett. {\bf 73}, 1762 (1994); Phys. Rev. Lett. {\bf 73}, 2809 (1994);
and references therein. } and right-handed neutrinos,  which no doubt opens
a window for new physics at the electroweak scale.
In fact, the former automatically
coincides with the requirement of the minimal supersymmetric
standard model (MSSM), and the latter becomes interesting
for particle physics, astrophysics and cosmology.
All other additional fields exist only above the GUT scale. They
consist of adjoint $45s$, exotic heavy 10s as well as some heavy singlets and
triplets of $\Delta_{48}(SU(3))$. Other higher representations of $SO(10)$
may also appear at the GUT scale. For simplicity, we will assume that
all 10s, 45s as well as other possible
higher representations of SO(10) are singlets of
$\Delta_{48}(SU(3))$. Obviously, when there are several fields
which belong to the same representations of $SO(10)\times \Delta_{48}(SU(3))$,
one has to introduce  additional quantum numbers to
distinguish those fields. Naive quantum numbers which can always be
assigned to complex fields are U(1) charges. For this simple reason,
$SO(10) \times \Delta_{48}(SU(3))$ is naturally extended to
$SO(10) \times \Delta_{48}(SU(3)) \times U(1)$. Obviously, such
a U(1) is family-independent. With this  symmetry, we can
construct interesting models to provide
an unified description for all the Yukawa and gauge couplings. Our focus in
this note is on fermion masses and mixings.  A realistic
model with an additional cyclic symmetry $C_{3}\times C_{2}$
is found to have the following superpotential.

 The superpotential which concerns interactions between different
fermions is given by
\begin{eqnarray}
W_{I} & = & \lambda_{H} \ \psi_{TDD}^{t} \sigma_{1} 10_{1} \psi_{TDD} +
\lambda_{\chi} \bar{\psi}_{TDD}\  \chi_{TDD}\  \psi_{TD3}
+ \lambda_{A} \bar{\psi}_{TD3}A_{TD}\psi_{D3}  \nonumber \\
& & +\lambda_{A3} \bar{\psi}_{D3}A_{D}\psi_{D2} + \lambda_{A2}
\bar{\psi}_{D2}A_{D}\psi_{D1} + \lambda_{A1} \bar{\psi}_{D1}A_{D}\hat{16}
\end{eqnarray}
and the superpotential for masses of the heavy $(16 + \bar{16})$ fermion pairs
is given by
\begin{eqnarray}
W_{M} & = & \lambda_{X} \bar{\psi}_{TDD}\  X_{TD}\  \psi_{TDD}
+ \lambda_{\tilde{A}}\bar{\psi}_{TD3} \tilde{A}_{D} \psi_{TD3} \nonumber \\
& & + S_{I}(\lambda_{S3}\bar{\psi}_{D3} \psi_{D3} + \lambda_{S2}
\bar{\psi}_{D2} \psi_{D2}) + \lambda_{S1} S_{P}\bar{\psi}_{D1} \psi_{D1}
\end{eqnarray}
where $\psi_{TDD}^{t}=(\psi_{TD1}^{t}, \psi_{TD2}^{t})$,
$\psi_{TDi}^{t}=(\psi_{Ti}^{t},\psi_{0i})$,
$\psi_{Ti}^{t}=(\psi_{1i}, \psi_{2i}, \psi_{3i})\  (i=1,2,3)$;
$\psi_{D\alpha}^{t}=(\psi_{\alpha}, \psi_{\alpha}')\ (\alpha =1,2,3)$
are heavy fermions  and
$ \chi_{TDD}^{t}=(\chi_{TD}, \chi)$, $\chi_{TD}=diag. (\chi_{T}, \chi_{0})$,
$\chi_{T} = diag. (\chi_{1}, \chi_{2}, \chi_{3})$,
$A_{TD}= diag. (A_{T}, A_{0})$; $A_{T}^{t}=( A_{x}, A_{z}, A_{u})$,
$A_{D}=diag.(A_{S}, A_{X})$, $\tilde{A}_{D}= diag. (A_{X}, A_{S})$,
$X_{TD}=diag. ( X_{T}, X_{0})$,
$X_{T}=diag. (X_{1}, X_{2}, X_{3})$  are heavy scalars.
Here the superscript `t' denotes transpose and `diag.' represents diagonal
matrix.  The cyclic symmetry $C_{3}$ acts on the
triplet fields labeled by the subscript `T' and the cyclic symmetry $C_{2}$
operates on the doublet fields labeled by the subscript `D'.
$\sigma_{1}$ is the  Pauli matrix acting on the doublet fields
$\psi_{TDD}^{t}=(\psi_{TD1}^{t}, \psi_{TD2}^{t})$.
Where the fields $\psi_{TD3} (\bar{\psi}_{TD3})$, and
$\psi_{D\alpha} (\bar{\psi}_{D\alpha})$ ($\alpha = 1,2,3$)
 belong to (16,$T_{1}$)($\bar{16}$, $\bar{T}_{1}$) representations
of $SO(10) \times \Delta_{48}(SU(3)) $; $\psi_{11}(\bar{\psi}_{11})$ and
$\psi_{12}(\bar{\psi}_{12})$ belong to $(16, T_{2})(\bar{16}, T_{2})$;
$\psi_{i1}(\bar{\psi}_{i1})$  and $\bar{\psi}_{i2}(\psi_{i2})$ ($i=2,3,0$)
belong to $(16,T_{3}) (\bar{16}, \bar{T}_{3})$. $X_{TD}$, $S_{P}$,
$S_{I}$,  $A_{S}$ and $A_{0}$ are singlets of $SO(10) \times
\Delta_{48}(SU(3))$. The 45s:   $A_{X}$, $A_{x}$,
$A_{z}$ and $A_{u}$ are singlets of $\Delta_{48}(SU(3))$. $(\chi_{1}, \chi_{2},
\chi_{3}, \chi_{0}, \chi)=(\bar{T}_{3}, T_{3}, \bar{T}_{1}, T_{2},
\bar{T}_{3})$. Once the triplet field $\chi$ develops an VEV only along the
third direction, i.e., $<\chi^{(3)}> \neq 0$, the resulting Yukawa coupling
matrices at the GUT scale will be automatically forced,
due to the special features of
$\Delta_{48}(SU(3))$, into an interesting texture structure with four
non-zero textures `33', `32', `22' and `12' which are characterized
by $\chi_{1}$, $\chi_{2}$, $\chi_{3}$, and $\chi_{0}$ respectively.
Here, the cyclic permutation symmetry $C_{3}\times C_{2}$ unifies
four textures into a single structure before symmetry breaking.
All the coupling constants $\lambda_{x}$ are real
due to CP symmetry and expected to be of order one.

  Considering the following symmetry breaking scenario motivated from the
succsessful prediction on the weak mixing angle $\sin\theta_{W}$
\begin{eqnarray}
& & SO(10) \times \Delta_{48}(SU(3)) \times U(1) \times C_{3}\times
C_{2} \stackrel{\bar{M}_{P}}{\rightarrow} SO(10)
\stackrel{v_{10}}{\rightarrow} SU(5) \nonumber \\
& & \stackrel{v_5\equiv M_{G}}{\rightarrow} U(1)_{Y}\times SU(2)_{L}
\times SU(3)_c  \stackrel{v_{1}, v_{2}}{\rightarrow} U(1)_{em}\times SU(3)_c
\end{eqnarray}
and the structure of the physical vacuum determined by the scalar fields
with $<\chi_{a}^{(i)}>=\bar{M}_{P}$ ($a=1,2,3,0;\ i=1,2,3$),
$<\chi^{(3)}>=\bar{M}_{P}$,
$<\chi^{(2)}>=<\chi^{(1)}>= 0$, $<A_{D}>=v_{10}$, $<A_{TD}>=v_{5}$,
$<S_{P}>=\bar{M}_{P}$, $<S_{I}>= v_{10}$, and
$(X_{1}, X_{2}, X_{3}, X_{0})=(v_{5}, v_{10}, v_{10}, \bar{M}_{P})$,
where the adjoint 45s develop VEVs along the directions:
$<A_{X}>=v_{10}\  diag. (2,\ 2,\ 2,\ 2,\ 2)\otimes \tau_{2}$;
$<A_{x}> =v_{5}\  diag. (2,\ 2,\ 2,\ 2,\ 2)\otimes \tau_{2}$; \
$<A_{z}> =v_{5}\  diag. (\frac{2}{3},\ \frac{2}{3},\ \frac{2}{3},\
-2,\ -2)\otimes \tau_{2}$;\  $<A_{u}>=v_{5}\
diag. (\frac{2}{3},\ \frac{2}{3},\ \frac{2}{3},\
 \frac{1}{3},\  \frac{1}{3})\otimes \tau_{2}$,
we then find that when integrating out the heavy fermion pairs, the Yukawa
coupling matrices  of the quarks and leptons at
the GUT scale $v_{5}\equiv M_{G}$ naturally have the following  symmetric
texture structure with zeros
\begin{equation}
\Gamma_{f}^{G} = \lambda_{G}
\left( \begin{array}{ccc}
0  &  \frac{1}{2}z_{f} \epsilon_{P}^{2}(1 + 2 \lambda_{W}^{2})  &   0   \\
\frac{1}{2}z_{f} \epsilon_{P}^{2}(1 + 2 \lambda_{W}^{2}) &  y_{f}
\epsilon_{G}^{2}(1 + 2 \lambda_{W}^{2}) e^{i\phi}
& \frac{1}{2}x_{f}\epsilon_{G}^{2}\sqrt{1 + 2 \lambda_{W}^{2}}  \\
0  &  \frac{1}{2}x_{f}\epsilon_{G}^{2}\sqrt{1 + 2 \lambda_{W}^{2}}  &  w_{f}
\end{array} \right)
\end{equation}
for\footnote{For neutrino masses and
mixings, we would like to discuss in detail elsewhere since it
needs a new mechanism to understand their small masses.}
 $f=u,d,e$. Where $\epsilon_{G}\equiv v_{5}/v_{10}$
and $\epsilon_{P}\equiv v_{5}/\bar{M}_{P}$. $x_{f}$, $y_{f}$, $z_{f}$ and
$w_{f}$ are the Clebsch-Gordon factors of $SO(10)$ and fixed
by the directions of symmetry breaking of the adjoint 45s:
$w_{u}=w_{d}=w_{e}=1$, $z_{u}=1$, $z_{d}=z_{e}= -27$,
$y_{u}=0$, $y_{d}=y_{e}/3=2/27$, $x_{u}= -7/9$,
$x_{d}= -5/27$, $x_{e}=1$.
$\phi$ is the physical CP phase\footnote{ We have rotated
away other possible phases by a phase redefinition of the fermion fields.}
arising from the VEVs \cite{SCPV}. Thus, for a given structure of the
physical vacuum, the above Yukawa Coupling matrices \footnote{We have
neglected small terms  arising from mixings between the chiral
fermion $16_{i}$ and
the heavy fermion pairs $\psi_{a} (\bar{\psi}_{a})$. Those small terms would
not change the numerical results by more than a few percent.}
$\Gamma_{f}^{G}$
only depend on two parameters
$\lambda_{G}$ (or $\lambda_{H}$) and $\lambda_{W}$ with
$\lambda_{G} \equiv 2\lambda_{W}^{2}\lambda_{H}/(1 + 2 \lambda_{W}^{2})$
and $ \lambda_{W} \equiv
(\lambda_{A1}\lambda_{A2}\lambda_{A3}\lambda_{A}\lambda_{\chi})
/(\lambda_{S1}\lambda_{S2}\lambda_{S3}\lambda_{\tilde{A}}\lambda_{X})$.

   In terms of an effective operator analysis, we have
\begin{eqnarray}
W_{33} & = & \frac{\lambda_{H}\lambda_{W}^{2}}{1 + 2 \lambda_{W}^{2}} \
16_{3} (\frac{A_{x}}{v_{5}})(\frac{v_{10}}{A_{X}})\ 10_{1}\
(\frac{v_{10}}{A_{X}}) (\frac{A_{x}}{v_{5}}) 16_{3}  \nonumber \\
W_{32} & = & \frac{\lambda_{H}\lambda_{W}^{2}}{\sqrt{1 + 2 \lambda_{W}^{2}}} \
\epsilon_{G}^{2}\ 16_{3} (\frac{A_{z}}{v_{5}})(\frac{v_{10}}{A_{X}})\ 10_{1}\
(\frac{v_{10}}{A_{X}}) (\frac{A_{z}}{v_{5}}) 16_{2}  \nonumber \\
W_{22} & = & \lambda_{H}\lambda_{W}^{2}\ \epsilon_{G}^{2}
\ 16_{2} (\frac{A_{u}}{v_{5}})(\frac{v_{10}}{A_{X}})\ 10_{1}\
(\frac{v_{10}}{A_{X}}) (\frac{A_{u}}{v_{5}}) 16_{2}  \\
W_{12} & = & \lambda_{H}\lambda_{W}^{2}\  \epsilon_{P}^{2} \
16_{1} (\frac{A_{X}}{v_{10}})^{3}\ 10_{1}\
(\frac{A_{X}}{v_{10}})^{3} 16_{2}  \nonumber
\end{eqnarray}
which is quite unique due to cyclic symmetry. Where the
uniqueness of the structure of the operator $W_{12}$
was first observed by \cite{OPERATOR}.
In addition to the common features resulting from most of the
predictive GUTs\cite{GJ,OPERATOR}, the present model exhibits its
new interesting features that lead to the fermion
Yukawa couplings only depending on a minimum set of
parameters. Unlike many other models in which  $W_{33}$ is assumed to be
a renormalizable interaction,  the Yukawa couplings
of all the quarks and leptons (both heavy and light) in the present model
are generated beneath the GUT scale when the
unified gauge symmetry is broken down
spontaneously. The hierarchy among the three families
arises from the mass hierarchy of the
heavy quark pairs $\psi_{ai} (\bar{\psi}_{ai})(i=1,2)$, which
originates from the hierarchic
VEVs developed from the four singlet fields $X_{a}$ ($a=1,2,3,0$).
Mass splittings between
the quarks and leptons as well as between the
up and down fermions result from the
Clebsch-Gordon factors of SO(10).
When scaling the GUT scale down to low energies, an additional
important effect for mass splitting
will arise from the Renormalization Group (RG) evolution.
Top-bottom splitting in the present model is mainly
attributed to the hierarchy of the VEVs $v_{1}$ and $v_{2}$ of
the two Higgs doublets.

    Our numerical predictions are presented in  table 1b
with the input parameters and their values given in table 1a.  Where the
RG effects have been considered following a standard
scheme\cite{THIRD,OPERATOR}, i.e.,
integrating the full two-loop RG equations from the GUT
scale down to the weak scale with taking
 $M_{SUSY} \simeq  M_{WEAK} \simeq M_{t} \simeq 180$GeV. From
the weak scale down to the lower energy scale, the fermion masses are obtained
by evolving to three loops from QCD and two loops in QED.

  {\bf Table 1a.}  Input parameters
and their values as the functions of the strong coupling $\alpha_{s}(M_{Z})$.
\\

\begin{tabular}{|c|c|c|c|c|c|c|}  \hline
$\alpha_{s}(M_{Z})$  &  $\lambda_{W}$  &  $\lambda_{G}$  &  $\phi$ &
$\epsilon_{G} \equiv v_{5}/v_{10}$ & $\epsilon_{P}\equiv v_{5}/\bar{M}_{P}$  &
$\tan\beta $   \\   \hline
0.110  &  0.95  &  0.64 $\lambda_{H}$  &  $73.4^{\circ}$  &  $2.75 \times
10^{-1}$  &   $ 0.92\times 10^{-2}$  & 51  \\
0.115  &  1.05  &  0.69 $\lambda_{H}$  &  $77.5^{\circ}$  &  $2.47 \times
10^{-1}$  &   $ 0.82\times 10^{-2}$  & 55  \\
0.120  &  1.09  &  0.71 $\lambda_{H}$  &  $81.5^{\circ}$  &  $2.27 \times
10^{-1}$  &   $ 0.75\times 10^{-2}$  & 58  \\   \hline
\end{tabular}
\\

{\bf  Table 1b.}  Output observables and their predicted values with the values
of the input parameters given in the table 1a.
\\

\begin{tabular}{|c|c|c|ccc|}   \hline
 Output   &     &   $\alpha_{s}(M_{Z})$  &  0.110  &   0.115
&  0.120   \\ \hline
$m_{b}(m_{b})$\ [GeV]  &  4.35   &  $M_{t}$\ [GeV]  &  165   &  176  &  185  \\
$m_{c}(m_{c})$\ [GeV]  &  1.22   &  $m_{s}$(1GeV)\ [MeV]
&  152   &  172 &  197  \\
$m_{\tau}$\ [GeV]  &  1.777   &  $m_{d}$(1GeV)\ [MeV]
&  6.5   &  7.2  &  8.0  \\
$m_{\mu}$\ [MeV]  &  105.6   &  $m_{u}$(1GeV)\ [MeV]
&  3.1   &  4.6  &  6.9  \\
$m_{e}$\ [MeV]  &  0.51   &  $|V_{cb}|\simeq A\lambda^{2}$
&  0.047   &  0.044  &  0.041  \\
$|V_{us}| \simeq \lambda$  & 0.220  &  $|\frac{V_{ub}}{V_{cb}}|\simeq \lambda
\sqrt{\rho^{2} + \eta^{2}}$  & 0.049  & 0.060  & 0.071    \\
  &    &  $|\frac{V_{td}}{V_{cb}}|\simeq \lambda
\sqrt{(1-\rho)^{2} + \eta^{2}}$  & 0.201  &  0.199  & 0.198  \\  \hline
\end{tabular}
\\

 From the table 1a, one sees that the two basic Yukawa
couplings $\lambda_{W}$ and $\lambda_{H}$
are truly close to the values of unity. The model has large
$\tan\beta$ solution with
$\tan\beta \equiv v_{2}/v_{1} \sim m_{t}/m_{b}$.
CP violation is near the maximum with a phase $\phi \sim 80^{\circ}$.
The vacuum structure between the GUT scale and Planck scale  has a hierarchic
structure $\epsilon_{G}\equiv v_{5}/v_{10} \sim \lambda = 0.22$ and
$\epsilon_{P}\equiv v_{5}/\bar{M}_{P} \sim \lambda^{3}$. Assuming
$(\bar{M}_{P}/M_{P})^{2} \simeq \alpha_{G} \simeq 1/24\sim \lambda^{2}$
(here $\alpha_{G}$ is
the unified gauge coupling, $M_{P}$ is the Planck mass), we have
\begin{eqnarray}
& & \bar{M}_{P}=2.5\times 10^{18} GeV, \nonumber \\
& & v_{10} \simeq (0.86 \pm 0.16) \times 10^{17} GeV, \\
& & v_{5}\equiv M_{G} \simeq (2.1 \pm 0.2) \times 10^{16} GeV \nonumber
\end{eqnarray}
where the resulting values for the GUT scale show a good agreement with the one
obtained from the gauge coupling unification. It implies that gravity would
become compatible with the unified gauge interaction at the family symmetry
breaking scale $\bar{M}_{P}$.
This is perhaps suggesting unity of gravity at the scale $\bar{M}_{P}$ which
may be regarded as a gravitational scale.

It is seen  from the table 1b that the resulting predictions on the
fermion masses and Cabbibo-Kobayashi-Maskawa (CKM)  mixing angles
show a remarkable  agreement with the experimental
data\cite{PDG,TOP,MASS}. The model also results a consistent prediction
for the $B^{0}$-$\bar{B}^{0}$ mixing and CP violation in kaon decays.

  It would be of interest to expand the above fermion Yukawa
coupling matrices $\Gamma_{f}^{G}$ in terms of the
parameter $\lambda=0.22$ (the Cabbibo angle),
which was  found \cite{WOLFENSTEIN} to be
very useful for expanding the CKM mixing matrix.
With the input values given in the table 1a, we find that
for $\alpha_{s}(M_{Z}) = 0.115$,  $\Gamma_{f}^{G}$ ($f=u,d,e $)
are given by
\begin{eqnarray}
& & \Gamma_{u}^{G} \simeq \lambda_{G} \left( \begin{array}{ccc}
0  & 0.95\lambda^{6}  & 0  \\
0.95\lambda^{6}  & 0  & -0.87\lambda^{2} \\
0  &  -0.87\lambda^{2}  &  1
\end{array}  \right); \  \Gamma_{d}^{G} \simeq \lambda_{G}
\left( \begin{array}{ccc}
0  & -1.24\lambda^{4}  & 0  \\
-1.24\lambda^{4}  & 1.36 \lambda^{3}\  e^{i0.86\pi/2}  & -0.95\lambda^{3} \\
0  &  -0.95\lambda^{3}  &  1
\end{array}  \right) \nonumber \\
& & \Gamma_{e}^{G} \simeq \lambda_{G} \left( \begin{array}{ccc}
0  & -1.24\lambda^{4}  & 0  \\
-1.24\lambda^{4}  & 0.90 \lambda^{2}\  e^{i0.86\pi/2}  & 1.13\lambda^{2} \\
0  &  1.13\lambda^{2}  &  1
\end{array}  \right) \ .
\end{eqnarray}

  It is amazing that nature has allowed us to make maximal predictions in terms
of a minimum set of parameters and to understand the low energy physics
from the Planck scale physics. It implies that the basic assumptions
on which we have based would reflect the right features of nature.
Nevertheless, nature does not yet expose its whole secrets
to us. The next step would be to  explore why nature chooses
that structure of the physical vacuum. It is expected that a deeper
understanding on nature will be arrived when unity of gravity is also
successfully realized.


\end{document}